\documentclass[letterpaper, 10pt, conference]{ieeeconf}  

\IEEEoverridecommandlockouts 

\usepackage{pifont}
\usepackage{macros}
\usepackage{multirow}
\usepackage{paralist,xspace}
\usepackage{tikz}
\newtheorem{theo}{Theorem}
\newtheorem{theorem}[theo]{Theorem}
\newtheorem{prob}{Theorem}
\newtheorem{problem}[prob]{Problem}

\newtheorem{defi}{Theorem}
\newtheorem{definition}[defi]{Definition}

\newcommand\learner{{learner}\xspace}
\newcommand\demonstrator{{demonstrator}\xspace}
\newcommand\verifier{{verifier}\xspace}

\renewcommand{\baselinestretch}{0.99}

\begin{document}

\title{\Large \bf Formal Policy Learning from Demonstrations \hadi{for Reachability Properties}}

\author{ Hadi Ravanbakhsh, Sriram Sankaranarayanan and Sanjit A. Seshia%
 \thanks{ Ravanbaksh and Seshia are affiliated with the University of California, Berkeley, USA. Sankaranarayanan is affiliated with the University of Colorado, Boulder, USA. }
}
\maketitle

\begin{abstract}
  
 We consider the problem of learning structured, closed-loop policies 
(feedback laws) from demonstrations in order to control under-actuated 
robotic systems, so that formal behavioral specifications such as 
reaching a target set of states are satisfied. Our approach uses a 
``counterexample-guided'' iterative loop that involves the interaction
between a policy learner, a demonstrator and a verifier. The learner 
is responsible for querying the demonstrator in order to obtain the 
training data to guide the construction of a policy candidate. 
This candidate is analyzed by the verifier and either accepted as 
correct, or rejected with a counterexample.  In the latter case, the 
counterexample is used to update the training data and further refine 
the policy.

The approach is instantiated using receding horizon model-predictive 
controllers (MPCs) as demonstrators. Rather than using regression 
to fit a policy to the demonstrator actions, we extend the MPC 
formulation with the gradient of the cost-to-go function evaluated at 
sample states in order to constrain the set of policies compatible 
with the behavior of the demonstrator. 
We demonstrate the successful application of the resulting policy 
learning schemes on two case studies
and we show how simple, formally-verified policies can be inferred 
starting from a complex and unverified nonlinear MPC implementations. 
As a further benefit, the policies are many orders of magnitude 
faster to implement when compared to the original MPCs.
\end{abstract}


\section{Introduction}~\label{sec:intro}


Policy learning (i.e., learning feedback control laws) 
is a fundamental problem in control theory and robotics,
with applications that include controlling under-actuated 
robotic systems and autonomous vehicles. The main challenge lies in 
designing a policy that provably achieves task
specifications such as eventually reaching a target set of states. 
In this paper, we present an automated approach to policy learning
with three goals in mind:
(a) compute policies that are guaranteed to satisfy a set of 
formal specifications, expressed in a suitable logic; 
(b) represent policies as a linear combination of a set of 
pre-defined basis functions which can include polynomials, trigonometric functions, or even user-provided functions,
and 
(c) compute policies efficiently, in real time. 
Finding policies that satisfy all three properties is not easy.
In this paper, we provide a partial solution to this problem in the form
of an automated method that learns from a demonstrator. Using two case studies, we show that complex controllers can
be replaced by much simpler policies that achieve all the three
desired goals stated above.

Our approach relies on a demonstrator component that can be queried
for a given starting state and demonstrates control inputs to achieve
the desired goals.  Specifically, we use nonlinear, receding horizon
model-predictive controllers (MPCs) as demonstrators. For a given
input, the MPC formulates a nonlinear optimization problem by
``unrolling'' a predictive model of the system to some time horizon
$T$. The constraints and the objectives will ensure that the behaviors
of the system over the time horizon will satisfy the properties of
interest, while optimizing some key performance metrics. A common
solution to this problem lies in training a policy that ``mimics'' the
input-output map of the MPC~\cite{levine2014learning,mordatch2014combining}. Instead, our approach is based on two new ideas. 
First, we extend the demonstrator to provide a range of
permissible control inputs for each state by using the gradient of the
MPC's cost-to-go function. This allows our search for a simple policy
to succeed more often. Second, we use a counterexample-guided approach
that iterates between querying the demonstrator to learn a candidate
policy compatible with the demonstrator query results thus far, and a
verifier that checks if the candidate verifier conforms to the
specifications, producing a counterexample upon failure. This minimizes
the number of demonstrator queries.

We demonstrate the applicability of our approach on two
case studies that could not be solved previously, comparing the new
policy learner with off the shelf supervised learning methods. The two
case studies involve (i) performing maneuvers on
a nonlinear ground vehicle model (illustrating the result in the
Webots\texttrademark~\cite{Michel2004} robotics simulator), and (ii) controlling a nonlinear model of a fixed
aircraft wing called the \emph{Caltech ducted
  fan}~\cite{jadbabaie2002control}.
We demonstrate in each case that our approach can
learn simple policies that satisfy all the desired requirements of
verification, simplicity, and fast computation. The resulting policies
are orders of magnitude faster to execute when compared to the
original MPCs from which they were learned.

\subsection{Related Work} 
Policy learning from
demonstrations is a fundamental problem in robotics, and the subject
of much recent work. Argall et al. provide a survey of various
learning from demonstration (LfD)
approaches~\cite{argall2009survey}. These approaches are primarily
distinguished by the nature of the demonstrators. For instance, the
demonstrator can be a human expert~\cite{khansari2017learning}, an
offline sample-based planning technique (e.g. Monte-Carlo Tree
Search~\cite{guo2014deep}), or an offline trajectory optimization
based technique~\cite{levine2014learning}. In particular, our approach
uses an offline receding horizon MPC to provide demonstrations.

An alternative to learning policies is to learn value (potential or
Lyapunov) functions. It is well known that systems that can be
controlled by relatively simple policies can require potential
functions that are complex and hard to learn. Thus, a vast majority of
approaches, including this paper, focus on policy learning.  However,
there have been approaches to learning value functions, including
Zhong et al.~\cite{zhong2013value}, Khansari-Zadeh et
al~\cite{khansari2017learning}, and our previous
work~\cite{ravanbakhsh2018learning}.  Notably, our previous work
queries demonstrators and uses counter-example learning in a similar
manner in order to learn potential (or control Lyapunov) function
described as an unknown polynomial of bounded degree. In contrast,
the approach of this paper
learns policies directly, and exploits the gradient of the
cost-to-go functions to make the demonstrator output a range of
control inputs. This allows for more policies to be retained at each
iterative step. At the same time, the fast convergence properties
established in our previous work are retained in our policy learning
framework.

Another important limitation in iterative policy learning is the lack
of an adversarial component that can actively identify and improve
wrong policies~\cite{ross2011reduction,he2012imitation,kahn2017plato}.
In contrast, our approach includes an adversarial verifier that
actively finds mistakes to fix the current policy. However, an
important drawback of doing so is our inability to learn on the actual
platform in real-time, unless the system can recover from violations
of the properties we are interested in. Currently, all the learning is
performed using mathematical models.  

The idea of generating controllers from rich temporal property
specifications underlies the field of formal synthesis. A variety of
recent approaches consider this problem, 
including discretization-based techniques that abstract the dynamics
to finite state machines and use automata-theoretic approaches to
synthesize
controllers~\cite{mazo2010pessoa,ozay2013computing,ravanbakhsh2014infinite,rungger2016scots},
formal parameter synthesis approaches that search for unknown
parameters so that the overall system satisfies its
specifications~\cite{yordanov2008parameter,taly2011synthesizing,jha-iccps10,abate2017sound},
deductive approaches that learn controllers and associated
certificates such as the Lyapunov
function~\cite{ravanbakhsh2015counter,el1994synthesis,tan2004searching,huang2015controller}.
Recent work~\cite{raman2015reactive,vazquez-adhs18}
presents approaches to controller synthesis for temporal logic
based on the 
paradigm of counterexample-guided
inductive synthesis (CEGIS)~\cite{solar2006combinatorial}.
Querying for demonstrations can be viewed as a way of actively
querying an oracle for positive examples, which is also done
in some CEGIS variants.
Our approach and these approaches are thus both instances of
the abstract framework of {\em oracle-guided inductive synthesis}~\cite{jha2017theory}.

\section{Background}~\label{sec:background} Let $\reals$ denote the
set of real numbers, $\reals^+$ denote non-negative real numbers, and
$\bools : \{ \mbox{true}, \mbox{false}\}$.  A column vector
$[x_1, x_2,..., x_n]^t$ is written as $\vx$.  We write $\vx \circ \vy$
to denote element-wise multiplication and $\vx \cdot \vy$ as the inner
product.  For a set $X$, let
$\vol(X)$ be the volume of $X$ and $\inter(X)$ be its interior.

In this paper, we consider the inference of a feedback function
(policy) for a given plant model and logical specification. 
Let $\X \subseteq \reals^n$ be a state-space whose elements are state
vectors written as $\vx$, and $\U \subseteq \reals^m$ denote a set of
control actions whose elements are control inputs written as
$\vu$. The plant model is described by a function
$\scr{F}: \X \times \U \rightarrow \reals^n$ that describes the right-hand
side of a differential equation for the states:
\[ \dot{\vx} = \scr{F}(\vx, \vu),\ \vx \in \X,\ \vu \in \U \,.\] For
technical reasons, $\scr{F}$ is assumed to be Lipschitz continuous in
$\vx$ and $\vu$.

\begin{definition}[Policy]
  A policy $\pi: \X \rightarrow \U$ maps each state $\vx \in \X$ to an
  action $\vu \in \U$.
\end{definition}

In this paper, we will consider policies $\pi$ that are Lipschitz
continuous over $\X$.  Given a plant, $\scr{F}$ and a policy $\pi$, a
closed loop system $\Psi(\scr{F},\pi)$ with initial state $\vx_0 \in X$
yields a trace (time trajectory) $\tr : \reals^+ \rightarrow \X$ that
satisfies
\[ \tr(0) = \vx_0,\ \mbox{and}\ (\forall\ t \geq 0)\  \dot{\tr}(t) = \plant(\tr(t), \pol(\tr(t))) \,. \]
Given a state-space $\X$, a specification $\varphi$ maps time
trajectories $\sigma: \reals^+ \rightarrow X$ to Boolean values
$\true/\false$, effectively specifying desirable vs. undesirable
trajectories, i.e,
$\varphi: (\reals^+ \rightarrow X) \rightarrow \bools$.  There are
many useful specification formalisms for time trajectories 
(e.g.,~\emph{Metric Temporal Logic}(MTL)~\cite{koymans1990specifying}).
While our approach is applicable to
the specification written in such a formalism, this work focuses
exclusively on reachability properties:
\begin{definition}[Reachability] Given a compact initial set 
$I \subset \X$ ($\tr(0) \in I$) 
and a compact goal set $G \subset \X$, the system must reach some 
state in $G$ :  $(\exists\ t \geq 0)\ \tr(t) \in G$.
\end{definition}
%

\begin{definition}[Policy Correctness] A policy $\pi$ is correct 
w.r.t. a specification $\varphi$, if for all traces $\tr$ of the system 
$\Psi(\plant,\pi)$, $\varphi(\tr)$ holds. For short, we write 
$\Psi(\plant, \pi) \models \varphi$.
\end{definition}

Given a plant $\scr{F}$ and specification $\varphi$, we consider the
policy synthesis problem in this paper.

\begin{problem}[Policy Synthesis Problem]\label{prob:main}
  Given a plant $\plant$ and specification $\varphi$, find a policy
  $\pol$ s.t. $\Psi(\plant, \pol) \models \varphi$.
\end{problem}

The policy synthesis problem asks for a controller that satisfies a
given formal specification.  This problem has been considered through
many formal synthesis approaches in the recent past~\cite{raman2015reactive,liu2013synthesis,huang2015controller,rungger2016scots,ozay2013computing}. In the subsequent section, we will describe the
specific setup considered in this paper.

\section{Formal Learning Framework}~\label{sec:framework}
\begin{figure}[t]
\vspace{0.2cm}
  \begin{center}
\begin{tikzpicture}
\matrix[every node/.style={rectangle, draw=black}, row sep=20pt, column sep=20pt]{
    \node(n0){\learner}; & \\
   \node(n1){\verifier }; & \node[fill=red!20](n3) {Output: \textsf{fail}}; \\
   \node(n2){\demonstrator};  & \node[fill=green!20](n4){Output: \textsf{succ.}}; \\    
};

\path[->, line width = 2pt] (n0) edge node[left]{$\pi_j$} (n1)  
(n1) edge node[left]{ $\vx_{j+1}$} (n2)
(n2) edge[in=180, out = 180] node[left]{$(\vx_{j+1}, U_{j+1})$}(n0)
(n0) edge node[right]{\textsc{No Candidate}} (n3)
(n1) edge node[right]{\ \textsf{SAT}} (n4);
\draw (n0.north)++(0,0.4cm) node[rectangle, draw=red, dashed]{$\{ (\vx_1,U_1), \ldots, (\vx_j, U_j) \}$ };
\end{tikzpicture}
\end{center}
\caption{Schematic diagram of the policy learning framework.}\label{Fig:formal-learning-framework}
\end{figure}
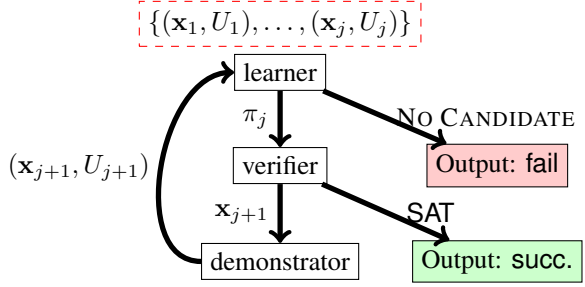

We propose a novel approach where the demonstrator provides a set of feasible feedback for a given state. 
Figure~\ref{Fig:formal-learning-framework} shows the three components
involved in the formal learning framework and their interactions.

The \learner is a core component that iteratively attempts to learn
a policy using the \demonstrator and \verifier components.
At the end, it either successfully outputs a policy
or outputs \textsc{fail}, indicating failure.

The learner works over a space of policies $\Pi$ fixed
a priori and  at each iteration, it maintains a finite \emph{sample} set
$O_i:\ \{ (\vx_1, U_1), (\vx_2, U_2),\ \ldots, (\vx_i, U_i) \}$,
with sample states $\vx_j \in \X$ and corresponding set of
control inputs $U_j \subseteq \U$. The set $O_i$ can be
viewed as a constraint over the set of policies in $\Pi$, using
the compatibility notion defined below:
\begin{definition}[Compatibility Condition]\label{def:learner-compatible}
  A policy set $\pi \in \Pi$ is compatible with $O_i:\ \{ (\vx_j, U_j), j = 1,\ldots,i \}$
  if and only if for all $(\vx_j, U_j) \in O_i$,
  $\pi(\vx_j) \in U_j$. 
\end{definition}

The \demonstrator $\dem$ inputs a state sample $\vx \in X$ and outputs a set
of control inputs $U \subseteq \U$. The \demonstrator outputs a
set of possible control inputs $U:\ \dem(\vx)$ that can be applied instantaneously at $\vx$. We require that the following correctness condition holds:
\begin{definition}[Demonstrator Correctness]\label{def:dem-correctness}
  For any policy $\pi$ if $(\forall \vx) \ \pi(\vx) \in \dem(\vx)$, then
  $\Psi(\plant, \pi) \models \varphi$.
\end{definition}
Therefore, we conclude that an incorrect policy $\pi$
``disagrees'' with the demonstrator output $\dem(\vx)$ for some state
$\vx$.
\begin{lemma}
Given a \emph{correct} \demonstrator $\dem$, a policy $\pi$ and a trace $\tr$ of  $\Psi(\plant, \pi)$ that violates the specification $\varphi$, there exists a time $t$ s.t. $\pi(\tr(t)) \notin \dem(\tr(t))$.
\end{lemma}


The \verifier inputs a policy $\pi$ and property $\varphi$, outputting
SAT or UNSAT. Here {SAT} signifies that the closed loop
$\Psi(\scr{F}, \pi)$ consisting of the plant and the current policy
satisfies the specification $\varphi$, and \textsc{UNSAT} signifies
that the closed loop fails the property. In the latter case, the
\verifier generates a trace $\tr: \reals^+ \mapsto \X$ of the closed
loop that violates the property.

\paragraph{Iteration:} At the start, the \learner is instantiated with
its initial policy space $\Pi$ (eg., all policies described that
are linear combinations of a set of given basis function) and
the initial sample set $O_0 = \emptyset$.

At the $i^{th}$ iteration, sample set
is denoted $O_i$. The following steps are carried out:
\begin{compactenum}
\item The \learner chooses a policy $\pi_i \in \Pi$ that is \emph{compatible} with
  $O_{i-1}$. Then, $\pi_i$ is fed to the \verifier.
\item The \verifier either accepts the policy as \textsc{SAT}, or provides a counterexample trace $\tr_i$.
\item Using the \demonstrator, a state $\vx_i : \tr_i(t)$ is found for which $\pi_i$ is not compatible with the \demonstrator. I.e. $\pi_i(\vx_i) \notin \dem(\vx_i)$.
\item The \learner updates $O_{i}:\ O_{i-1} \cup \{ (\vx_i, \dem(\vx_i)) \}$.
\end{compactenum}

\begin{theorem}
  If the formal learning framework terminates with \textsc{success}, then
  we obtain a policy $\pi$ such that $\Psi(\plant, \pi)$ satisfies the
  desired property $\varphi$.
\end{theorem}

\section{Realizing the Oracles}~\label{sec:impl}

For simplicity, we will first start by describing how a demonstrator
can be realized.

\paragraph{Demonstrator:} Given a state $\vx \in \X$, the demonstrator
should output a set of control inputs $U \subseteq \U$ such that each
input $\vu \in U$ can be applied at $\vx$ without compromising the desired
property $\varphi$.
We will focus on describing a demonstrator for reachability properties
for now. A more general framework is left for future work.

Let $G$ be a set of \emph{goal states} that we wish to reach. We will
assume the following properties of our demonstrator:

\noindent\textbf{D1:} The demonstrator has an (inbuilt) policy $\pi^*$ that can
  ensure that starting from any $\vx \in \X$, the resulting trace
  $\sigma(t)$ reaches $G$ in finite time.

  Given any point $\vx$, we assume that $\pi^*(\vx)$ can be computed for any $\vx \in \X$. However, such a computation can be expensive and we do not know $\pi^*$ in a closed form.

\noindent\textbf{D2:} The demonstrator's correctness is certified by a
  smooth \emph{Lyapunov-like} (value) function $V$ such that
  \begin{compactenum}[(a)]
  \item for all $\vx \in \X \setminus \inter(G)$, $V(\vx) \geq 0$,
  \item $V$ is radially unbounded, and
  \item For all $\vx \in \X \setminus \inter(G)$, with $\vu = \pi(\vx)$,
  \[ \nabla_x V \cdot \scr{F}(\vx, \vu) \leq - \epsilon \,.\]
  Once again, we assume that $V$ is not available in a closed form but
  that $V(\vx)$ and $\nabla V(\vx)$ can be computed for each $\vx \in \X$.
\end{compactenum}

 Consider a demonstrator with a policy $\pi^*$ satisfying \textbf{D2}.
\begin{lemma}
  Starting from any state $\vx \in \X \setminus \inter(G)$, the closed
  loop trajectory of $\Psi(\plant, \pi^*)$ eventually reaches
  $\inter(G)$ in finite time.
\end{lemma}

Let $\lambda \in (0,1]$ be a chosen constant. We will denote
$\pi^*(\vx)$ by $\vu^*$.  For any $\vx \in \X \setminus \inter(G)$, we
define the set $U_{\lambda}(\vx) \subseteq \U$ as satisfying:
\begin{equation}\label{eq:U-extraction}
  U_{\lambda}(\vx):\ \{ \vu \in \U\ |\ (\nabla V) \scr{F}(\vx, \vu) \leq \lambda (\nabla V) \scr{F}(\vx, \vu^*) \}\,.
  \end{equation}
\begin{theorem}
  For any $\lambda \in (0,1]$ and any policy $\pi$ such that
  $ (\forall\ \vx \in \X \setminus \inter(G))\ \pi(\vx) \in
  U_{\lambda}(\vx)$, the closed loop model $\Psi(\plant,\pi)$ will
  satisfy the reachability property for $G$ (demonstrator correctness).
\end{theorem}

\paragraph{MPC Based Demonstrator:} 
We will now briefly consider how to implement a demonstrator for a
reachability property and extract a suitable proof $V$. To do so, we
will use a standard receding horizon MPC trajectory optimization
scheme that discretizes the plant dynamics using a discretization step
$\delta > 0$ and a terminal cost function that is chosen so that the
resulting MPC stablizes the plant $\scr{F}$ to a point
$\vx^* \in \inter(G)$.  Let $\hat{F}(\vx,\vu)$ be a discretization of
$\scr{F}$ for the time-step $\delta$.  This discretization
approximates $\scr{F}$ and is derived using Euler or Runge-Kutta
scheme. We define $V^*(\vx)$ as the optimal value of the following
problem:
\begin{equation}\label{eqn:opt-demonstrator}
 \begin{array}{ll}
     \underset{\vu(0),\ldots,\vu(N\delta-\delta)}{\mathop{\min}} & \sum\limits_{j=0}^{N-1}Q(\vu(j\delta),\vx(j\delta))+H(\vx(N\delta))\\
     				   \mathsf{s.t.} & \vx(0) = \vx \\
                        & \vx((j+1)\delta) = \hat{F}(\vx(j\delta), \vu(j\delta)) \\ & \ j = 0,\ldots, N-1 \,.
                       \end{array}
                     \end{equation}
Likewise, $\pi^*(\vx)$ is the optimal value  for $\vu(0)$ in~\eqref{eqn:opt-demonstrator}. 
 Under some well-known (and well-studied) conditions, the MPC scheme
 stabilizes the closed-loop dynamics to $\vx^*$ with the optimal cost to go $V^*$ as the
 desired Lyapunov function~\cite{Mayne+Others/2000/Constrained,Jadbabaie+Hauser/2005/Stability}.  We consider the following strategy for 
 a demonstrator:
 \begin{enumerate}
 \item We design an MPC controller with proper cost function (usually 
 $Q$ and $H$ are positive outside $\inter(G)$ and radially unbounded).
 \item We adjust the cost function by trial and error until
   the demonstrator works well on sampled initial states $\vx \in \X$
   and the cost decreases strictly along each of the resulting
   trajectories.
 \item The gradient $\grad V^*$ can be estimated for a given $\vx$ by
   using the KKT conditions for the optimization
   problem~\eqref{eqn:opt-demonstrator} or using a numerical scheme that
   estimates the gradient by sampling around $\vx$. We also note that some
   methods like iLQR~\cite{li2004iterative} provide local $V^*$ in closed form (and thus $\grad V^*$) along with the solution $\pi^*$.
 \end{enumerate}

\paragraph{Verifier:}
Given a policy $\pi$ and a property $\varphi$, the \verifier checks
whether the closed loop $\Psi(\plant, \pi)$ satisfies the property
$\varphi$ and if not, produces a counterexample trace.  The problem
of verifying non-trivial properties of nonlinear systems is
undecidable. Therefore a perfect \verifier is not feasible.

In this section, we review two main sets of solutions: (a) A \verifier
that attempt to approximately solve the verification problems using
decision procedures, as described in our earlier
work~\cite{ravanbakhsh2018learning}.  Such a \verifier concludes that the
system satisfies the property or is likely buggy. It produces an
\emph{abstract counterexample} that does not need to correspond to a real
trace of $\Psi(\plant,\pi)$ but could nevertheless be used in the
learning loop (see~\cite{ravanbakhsh2018learning} for
further details); or alternatively (b) a \emph{falsifier} that tests
$\Psi(\plant,\pi)$ for a large number of (carefully chosen) initial
states $\vx \in \X$, concluding either a real counterexample or that
the system likely satisfies the property.

In this paper, we consider falsifiers that ``invert'' the optimization
problem from Eq.~\eqref{eqn:opt-demonstrator}, as a  search heuristic for a
counterexample to the property of reaching the goal $G$.
\begin{equation}\label{eqn:opt-falsifier}
  \begin{array}{ll}
     \underset{{\vx(0) }}{\mathbf{\max}}\ & \sum\limits_{j=0}^{N-1} (Q(\vu(j\delta), \vx(j\delta))) + H(\vx(N\delta)) \\
    \mathsf{s.t.} & \vx((j+1)\delta) = \hat{F}(\vx(j\delta), \vu(j\delta)) \\
    & \vu(j\delta) = \pi(\vx(j\delta)) \,. \\
                       \end{array}
 \end{equation}
 However, the optimization problem is over the unknown initial state
 $\vx(0)$ with the control values $\vu(j\delta)$ fixed by the policy
 $\pi$ to be falsified. We attempt to solve this problem by using a
 combination of random choice of various $\vx(0)$ and a second order
 gradient descent search. Whereas this is not guaranteed to find a
 falsifying input, it often does within a few iterations,
 outperforming random simulations. On the other hand, if $M \geq 10^6$
 (or a suitably large number) of trials do not yield a falsification,
 we declare that the policy likely satisfies the property.

\paragraph{Witness State Generation:} %
Having found a counterexample trace $\tr$, we still need to find a
\emph{state} $\vx_{i} = \tr(t)$ of the trace to return back to the
demonstrator. Ideally, we choose a state $\vx:\ \tr(t)$ in the trace
at time $t$ such that $\pol_i(\vx) \notin U_{\lambda}(\vx)$, wherein
$U_{\lambda}(\vx)$ is the set returned by the \demonstrator (derived from the Lyapunov conditions) for input
$\vx$ (Cf. Eq.~\eqref{eq:U-extraction}).  
For this purpose, we
discretize the time (using small enough time-step $\delta$). 
At each discrete time $t$,
if the policy $\pi$ is compatible with the \demonstrator at $\tr(t)$,
we increment $t$.

\section{Learner}
Given a set of observations $O_i$, the learner finds a policy that is compatible with the observations. Formally, the policy is parameterized by $\param$ and the policy space $\Pi$ is represented by the parameter space $\Param$. The learner wishes to find $\param$ s.t.
\begin{equation}\label{eq:learner}
(\forall (\vx_j, U_j) \in O_i) \ \pol_{\param}(\vx_j) \in U_j \,.
\end{equation}
This will be posed as a  system of constraints over $\param$. Also, let $\Param_i \subseteq \Param$ be set of all such $\param$.


%

Let $\scr{V}$ be a finite set of basis functions $\{v_1,\ldots,v_K\}$ and
$\pi_{\param}$ be linear combinations of these functions:
\[\pi_{\param}(\vx) : \sum_{k=1}^K \param_k v_k(\vx) \,.\]
First, $\pol_\param$ is a linear function of $\param$.
We will now derive the constraints for the
compatibility condition (Def.~\ref{def:learner-compatible}),
given a sample set $O_i: \{ (\vx_1, U_1), \ldots, (\vx_i, U_i) \}$.

We will assume that each set $U_j$ is  a polyhedron
\[
U_j : \{\vu \ | \ A_{j} \vu \leq \vb_{j} \} \,.
\]

Therefore, in Eq.~\eqref{eq:learner}, $\pol_{\param}(\vx_j) \in U_j$
can be replaced with
$A_{j} (\sum_{k=1}^K \param_k v_k(\vx_j) ) \leq \vb_{j}$. Since
$\vx_j$ is known, the compatibility conditions yield a polyhedron over
$\param$.

\begin{theorem}\label{thm:linear-model}
  The compatibility conditions $\Param_i$, given a sample set $O_i$,
  form a linear feasibility problem.
\end{theorem}


Using ideas from Ravanbakhsh et al~\cite{ravanbakhsh2018learning},
we show that the entire formal learning algorithm terminates in
polynomial time, if the learner selects the new parameters
$\param \in \Param_i$ at each iteration, carefully.


By Theorem.~\ref{thm:linear-model}, $\Param_i$ for iteration $i$ would
be a polyhedron.

We consider two realistic assumptions:
\begin{enumerate}
	\item $\Param_0$ is a compact set, where $\Param_0 \subseteq [-\Delta, \Delta]^K$, $\Delta > 0$ is an arbitrarily large constant.
	\item The formal learning algorithm terminates whenever a $K$-ball of radius $\delta$ does not fit inside $\Param_i$ for some arbitrarily small $\delta > 0$ (not when $\Param_i = \emptyset$).
\end{enumerate}

We design the learner in the following way. In the learning process, $O_i$ implicitly defines $\Param_i$. Given $\Param_i$, let the learner return the center of the maximum volume ellipsoid (MVE) inside $\Param_i$. The problem of finding the MVE is equivalent to solving a SDP problem\cite{vandenberghe1998determinant}.

\begin{theorem}\label{thm:convergence}
	If the learner returns the center of MVE inside $\Param_i$ at each iteration, the formal learning algorithm terminates in $\frac{K (\log(\Delta) - \log(\delta))}{- \log\left(1 - \frac{1}{K}\right) } = O(K)$ iterations.
\end{theorem}

This theorem addresses an important issue in statistical machine
learning. If the model is not precise enough to capture a feasible
policy, the learning procedure terminates. And one can use a more
complicated model with a larger set of features (basis functions). In
other words, one can start from a simple model with smaller number of
parameters and iteratively add new basis functions.

We will now illustrate the results of policy learning for linear
combinations of basis functions using two case studies.
The \demonstrator is implemented by an MPC with quadratic cost
functions:
\[Q = [\vu^t \ \vx^t] \ \mbox{diag}(Q') \ [\vu^t \ \vx^t]^t \,, \ H = \vx^t \ \mbox{diag}(H') \ \vx \,.\]
A second order method is used to solve
Eq.~\eqref{eqn:opt-demonstrator}. For the falsifier, we use one
million ($10^6$) simulations from randomly generated initial states, interspersed with $10^3$ iterations of the adversarial falsifier chosen by solving eq.~\eqref{eqn:opt-falsifier}. If a
counterexample is found, it is reported. Otherwise, the policy is
declared \emph{likely correct}. Also, we use the demonstrator to randomly generate a single trace and initialize the dataset with demonstrations along that trace.

\begin{figure}[t]
\vspace{0.2cm}
\begin{center}
	\includegraphics[width=0.45\textwidth]{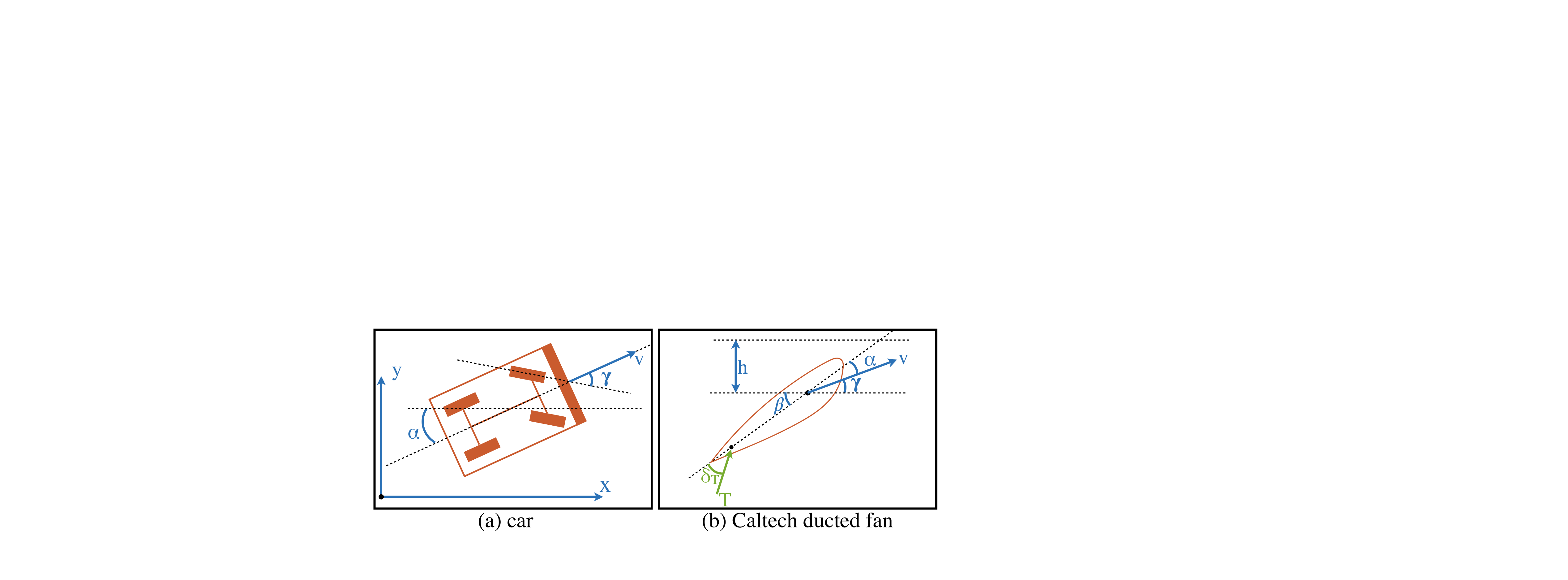}
\end{center}
\caption{Schematic View of the Case-Study I (a) and II (b)}\label{fig:schematics} 
\end{figure}

\paragraph{Case-Study I (Car):} 
A car with two axles is modeled by state variables $\vx^t = [x \ y \ v \ \alpha \ \beta]$, where $\beta = \tan(\gamma)$ and $\gamma$ is the degree between the front and back axles (Fig.~\ref{fig:schematics}(b)). The dynamics are defined as follows:
\[
\dot{x} = v \cos(\alpha) \,, \dot{y} = v \sin(\alpha) \,, \dot{v} = u_1 \,, \dot{\alpha} = \frac{v}{b}\beta \,, \dot{\beta} = u_2 \,,
\]
where $b = 3$. Also, $u_1 \in [-1, 1]$ and $u_2 \in [-3, 3]$ are inputs.

The goal is to follow a reference curve (the road), by controlling the
lateral deviation from the midpoint of the reference, at constant
speed $v_0 = 10 \text{m/s}$. For convenience the $y$-axis always coincides with
this lateral deviation. The state variable for $x$ is ignored in our
model and velocity $v$ is taken relative to $v_0$ (i.e, $v := v - v_0$).
The goal set is $G : (y, v, \alpha, \beta) \in [-0.1, 0.1]^4$ and the
initial set $I : (y, v) \in [-2, 2]^2 \times (\alpha, \beta) \in [-1, 1]^2$. For the MPC, the cost
functions are defined by $Q' : [1\ 1\ 9 \ 9 \ 1 \ 1]$ and $H' : [90
  \ 90 \ 10 \ 10]$, $\delta = 0.2$, $N = 10$. We test that the MPC
can solve the control problem for many random initial states, whereas a LQR controller with the same cost
function fails: I.e, starting from $I$, the goal $G$ is not reached by
some of the executions of this controller. 
Since there are two inputs, we use two different parameterizations
$\pol_\param : [\pol^1_{\param_1}, \pol^2_{\param_2}]$, where
$\pol^1_{\theta_1}$ ($\pol^2_{\theta_2}$) is used to learn $u_{1}$
($u_{2}$). We consider each input to be an affine combination of
states (affine policy). Our approach successfully finds an affine policy with
only $16$ demonstrations.

To study scalability of the method, we consider varying number of cars
which do not directly interact. Our goal is for each lateral deviation
to converge to a narrow range using a single ``centralized'' policy to
control all of the cars at the same time. For $l$ cars, there are $2l$ inputs, each of which is an affine (linear)
feedback with $4l+1$ ($4l$) terms. The results for up to $4$ cars is
shown in Table~\ref{tab:bicycle}.

Results indicates that the method converges much faster when the
policy is linear as opposed to being affine. This suggests that
selecting basis functions can significantly affect the
performance. Nevertheless, the termination is guaranteed and the
method is scalable to higher dimensional problems as the complexity is
polynomial in the number of states. After all, we are using local
search for falsifier and demonstrator and using SDP solvers to
implement the learner.

We note that falsification is quite fast for most of iterations and it
takes significant time only at the final iteration, where the model
is most likely correct. Moreover, the witness
generation is the most expensive computation especially for larger
problems. However, such active search for witnesses helps to generate
useful data and guarantees convergence of the algorithm.

\begin{figure}[t]
\begin{center}
	\includegraphics[width=0.45\textwidth]{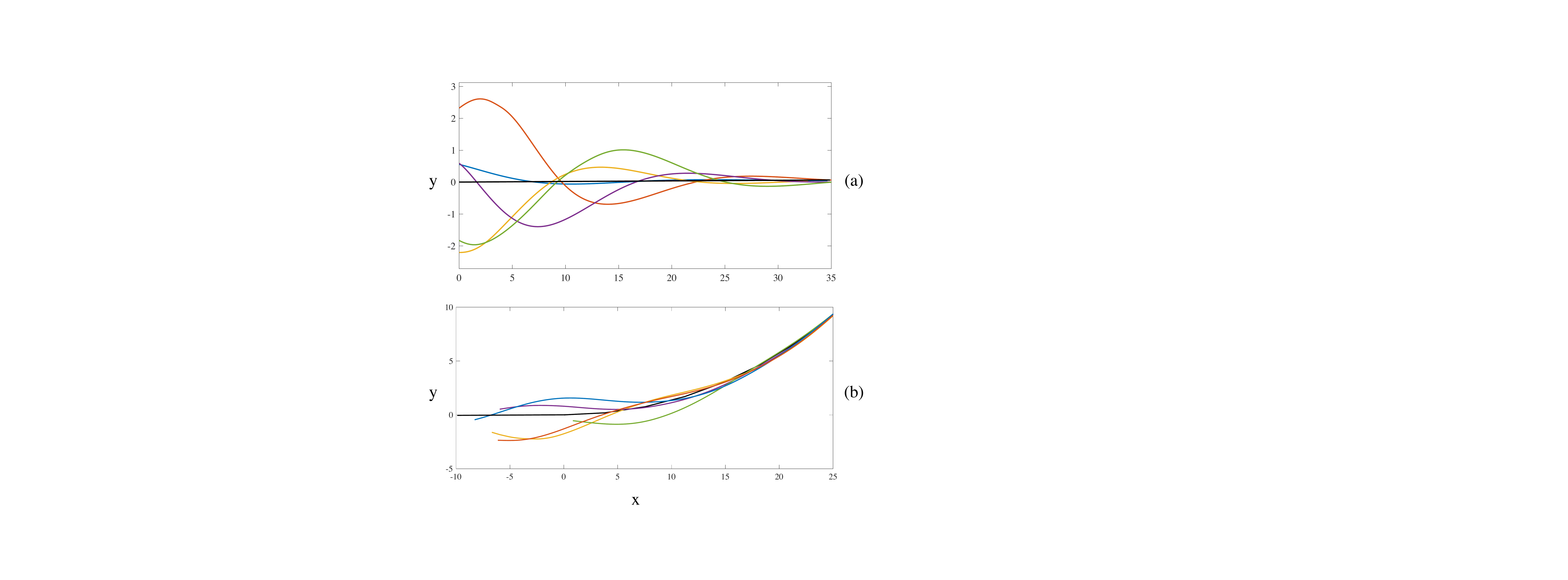}
\end{center}
\caption{Simulation Results in Webots for Two Road Segments: (a) a straight road, and (b) a curved road. The center of the road is shown with a black line and traces are shown with colored lines starting from different states.}\label{fig:webots} 
\end{figure}

We implemented the controller for a car in the
Webots\texttrademark~\cite{Michel2004} simulator.  The cars in Webots have a map
for the road and simulate GPS-based localization with internal sensors
for heading, steering angle and velocity. The car model in Webots is
much more complicated when compared to our simple bicycle
model. However, we use the bicycle model to design the policy.  The
simulation traces for some random initial states are shown in
Fig.~\ref{fig:webots} demonstrate that the learned controller is robust enough to compensate the model
mismatches. The simulations are shown for a straight as well as a
curved road segment.

\begin{table}[t]
\vspace{0.2cm}
	\caption{Results for Case-study I}
\begin{center}
	\begin{tabular}{||c|c||c|c|c||c||}
\hline
 $\#$Cars & K & $\#$Itr. & Wit. Gen. T. & Fals. T. & Total T. \\
 \hline
\multirow{2}{*}{1} & 8  & 2 & 3 & 59 &  66 \\
                   & 10 & 10 & 22 & 40 & 65\\
\hline
\multirow{2}{*}{2} & 32 & 13 & 115 & 98 & 233\\
                   & 36 & 33 & 423 & 55 & 492\\
\hline
\multirow{2}{*}{3} & 72 & 16 & 274 & 364 & 680 \\
				   & 78 & 44 & 1903 & 141 & 2084 \\
\hline
\multirow{2}{*}{4} & 128 & 47 & 1360 & 257 & 1697\\
				   & 136 & 62 & 6766 & 187 & 7087\\
\hline
\end{tabular}
\end{center}
\label{tab:bicycle}
\hadi{$K$ is the number of model parameters. }$\#$Itr. is the number of iterations. Total T. is the total computation time in seconds on a Mac Book Pro with up to 4 GHz Intel Core i7 processor.\\
\end{table}

\begin{figure*}[t]
\vspace{0.2cm}
\begin{center}
	\includegraphics[width=0.9\textwidth]{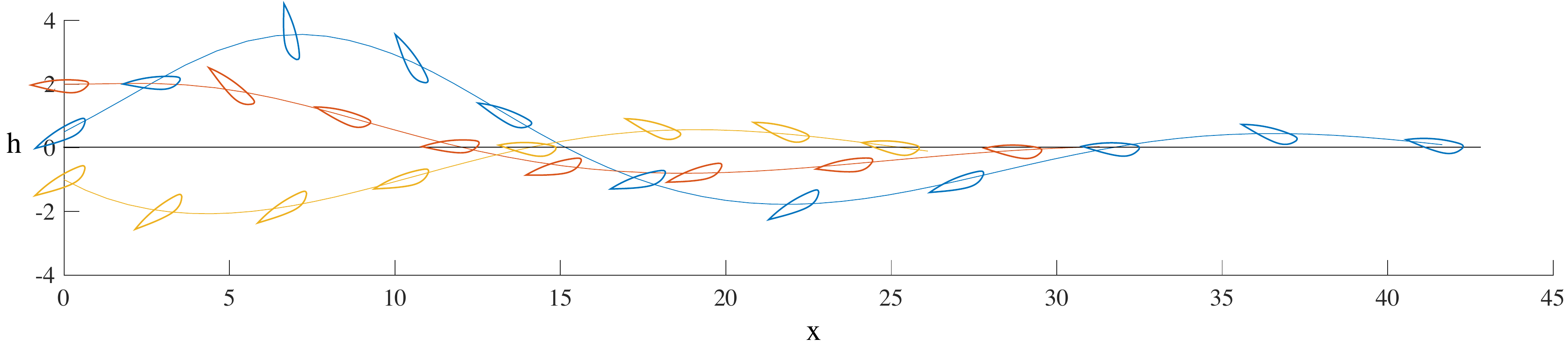}
\end{center}
\caption{Three execution traces of Case-Study II for the learned policy. $h$ is the height and $x$ is the horizontal position (initially, $x = 0$). The wings demonstrate the value of $\beta$ for some time instances. Initial state are as follows. blue:[1 0.5 0.5 0.5 0.5]$^t$, red:[0 0 0 0 2]$^t$, and yellow:[-1 -0.5 0.5 0.5 -1]$^t$.}\label{fig:sim-ducted-fan} 
\end{figure*}

\paragraph{Case-Study II (Caltech Ducted Fan):} We
consider an example from Jadbabaie et al.~\cite{jadbabaie2002control},
where authors develop a mathematical model for the Caltech ducted fan. The
state $\vx^t = [v \ \gamma \ \beta \ \dot{\beta} \ h]$, consists of speed $v$, angle of velocity $\gamma$, angle
of the ducted fan $\beta$, angular velocity $\dot{\beta}$, and height
$h$. Fig.~\ref{fig:schematics}(b) shows a schematic view of the
ducted fan.

The problem here is to stabilize the wing to move forward. The dynamics are:
\begin{align*}
	m \dot{v} = & -D(v, \alpha) - W \sin(\gamma) + u \cos(\alpha + \delta_u)  \\
	m v \dot{\gamma} = & L(v, \alpha) - W \cos(\gamma) + u \sin(\alpha + \delta_u) \\
	J \ddot{\beta} = & M(v, \alpha) - u l_T \sin(\delta_u)\\
	\dot{h} = & v \sin(\gamma) \,,
\end{align*}
	
where $\alpha = \beta - \gamma$, $u$ is the thrust force, and
$\delta_u$ in the angle for direction of the thrust
($\vu^t = [u \ \delta_u]$)
(Cf.~\cite{jadbabaie2002control} for a full description).

The model has a steady state $\vx^{*t} = [v_0 \ 0 \ \beta_0 \ 0 \ 0]$, where
$v_0 = 6$ and $\beta_0 = 0.177$ (for input $\vu^{*t} = [3.2 \ -0.138]$),
for which the ducted fan steadily moves forward.  The fact that
dynamics are not affine in control is problematic as we need each $U_j$ to be a polyhedron. To get around this, we
use $u_{s} : u \ {\sin}(\delta_u)$ and $u_{c} : u \ {\cos}(\delta_u)$ as our
inputs, and rewrite the equations. The dynamics of the new system are
affine in control and $U_j$ would be a polyhedron. By setting $\vx^*$ and
$\vu^*$ as the origins and scaling down $v$, by $2.5$ times
\[ 
\vx := [0.4 \ 1 \ 1 \ 1 \ 1]^t \circ (\vx - \vx^*) \ , \ \vu := \vu - \vu^* \,,
\]
the goal is to reach $G : [-0.2, 0.2]^5$ from $I: [-0.5, 0.5]^5$.
For the MPC, we use $Q': [0.02 \ 0.02\ 1 \ 1\  1\  1\  1]$, $H':[15\  15\  15\  15\  15]$, $\delta = 0.3$, $N=15$.
Let $\scr{T} : \{\tau_1(\vx),...,\tau_{K'}(\vx)\}$ be set of terms defined over $\vx$. Basis functions $v_k(\vx):\ \scr{T}^{\alpha_k}$ ($\pol_{\param} : \sum_k \param_k \ \scr{T}^{\alpha_k}$) correspond to monomials of terms in $\scr{T}$, wherein $\alpha_k \in \mathbb{N}^{K'}$ is a vector of natural number powers such that $|\alpha_k|_1 \leq d$ for some degree bound $d > 0$.
Both inputs are parameterized with terms $\scr{T}:\{v, \gamma, \beta, \dot{\beta}, h, \sin(\beta), \cos(\beta)\}$ and degree $d$ is $2$. Using the formal learning algorithm with $\lambda = 0.1$, proper parameters are found with $46$ demonstrations. Several traces of $\Psi(\plant, \pol)$ for the learned policy $\pol$ is shown in Fig.~\ref{fig:sim-ducted-fan}.

The results suggest that our framework can efficiently yield reliable solution to reachability problems, given appropriate basis. Nevertheless, if the bases are not rich enough, the method declares failure after few iterations. For example, when $\scr{T}:\{v, \gamma, \beta, \dot{\beta}, h\}$, the method terminates in $20$ iterations without any solution.

\paragraph{Comparison with Linear Regression:}
We consider a simple supervised learning algorithm in which the demonstrator generates optimal input $\vu_i$ for a given state $\vx_i$. Using the demonstrator we generate optimal traces starting from $M$ random initial states. Then, the states in the optimal traces (and their corresponding inputs) are added to the training data.
Having a dataset $\{(\vx_1, \vu_1), \ldots, (\vx_j, \vu_j)\}$, one wishes to find a policy with low error (at least) on the training dataset. In a typical statistical learning procedure, one would minimize the error between the policy and demonstrations:
\[\min_\theta \sum_{i=1}^j (\pi_\theta(\vx_i)-\vu_i)^2\,.\]
For both case studies, we tried this statistical learning approach with different training data sizes: $10 < M < 1000$. 
In all experiments the falsifier found a trace where the learned policy fails.
We believe this notion of error used in regression is merely a heuristic and may not be relevant. For example, we noticed that because of input saturation, many states in the training data have exactly the same corresponding control inputs in the dataset and this prevents a simple linear regression to succeed.

\section{Conclusion}
We have presented a policy learning approach that combines
learning from demonstrations with formal verification,
and demonstrated its effectiveness on two case studies. 
We showed cases where naive supervised learning fails due to its 
simplicity, whereas our method can solve the problem. Our future 
work will consider nonlinear models such as deep neural networks
as well as extensions to support a richer set of properties 
beyond reachability.

\hadi{
\section*{Acknowledgments}
This work was funded in part by NSF under award numbers 
SHF 1527075, CPS 1646556, and CPS 1545126, by the DARPA Assured
Autonomy grant, and by Berkeley Deep Drive. 
}

\bibliographystyle{plain}
\bibliography{references}

\end{document}